\documentclass[%
aip,
jcp,%
amsmath,amssymb,
reprint,%
]{revtex4-1}
\usepackage[latin1]{inputenc}

\usepackage{amsmath}
\usepackage{amsfonts}
\usepackage{amssymb,bm}
\usepackage{graphicx}
\usepackage{physics}
\usepackage{upgreek}
\usepackage{xcolor}
\usepackage[outercaption]{sidecap}   
\usepackage{txfonts}

\DeclareMathAlphabet{\pazocal}{OMS}{zplm}{m}{n}
\newcommand{\pP}{\ensuremath{\pazocal{P}}}
\newcommand{\pQ}{\ensuremath{\pazocal{Q}}}
\newcommand{\pL}{\ensuremath{\pazocal{L}}}
\newcommand{\pK}{\ensuremath{\pazocal{K}}}

\newcommand{\pD}{\ensuremath{\pazocal{D}}}

\newcommand{\el}{\ensuremath{\mathrm{e}}}

\newcommand{\sing}{\ensuremath{\mathrm{S}}}
\newcommand{\trip}{\ensuremath{\mathrm{T}}}

\renewcommand{\op}[1]{\ensuremath{\hat{#1}}}

\newcommand{\kT}{k_\trip^{\mathrm{r}}}
\newcommand{\kS}{k_\sing^{\mathrm{r}}}
\newcommand{\kD}[2]{k_{#1 #2}^{\mathrm{d}}}
\newcommand{\highlight}[1]{{#1}}
\usepackage{filecontents}

\begin{document}
	
\title{Radical pair intersystem crossing: quantum dynamics or incoherent kinetics?}
\author{Thomas P. Fay}
\email{thomas.fay@chem.ox.ac.uk}
\affiliation{Department of Chemistry, University of Oxford, Physical and Theoretical Chemistry Laboratory, South Parks Road, Oxford, OX1 3QZ, UK}
\author{David E. Manolopoulos}
\affiliation{Department of Chemistry, University of Oxford, Physical and Theoretical Chemistry Laboratory, South Parks Road, Oxford, OX1 3QZ, UK}

\begin{abstract}
Magnetic field effects on radical pair reactions arise due to the interplay of coherent electron spin dynamics and spin relaxation effects, a rigorous treatment of which requires the solution of the Liouville-von Neumann equation. However, it is often found that simple incoherent kinetic models of the radical pair singlet-triplet intersystem crossing provide an acceptable description of experimental measurements. In this paper we outline the theoretical basis for this incoherent kinetic description, elucidating its connection to exact quantum mechanics. 
We show in particular how the finite lifetime of the radical pair spin states, as well as any additional spin-state dephasing, leads to incoherent intersystem crossing. We arrive at simple expressions for the radical pair spin state interconversion rates to which the functional form proposed recently by Steiner {\em et al.} [J. Phys. Chem. C \textbf{122}, 11701 (2018)] can be regarded as an approximation. We also test the kinetic master equation against exact quantum dynamical simulations for a model radical pair and for a series of $\text{PTZ}^{\bullet+}\text{--Ph}_\text{n}\text{--PDI}^{\bullet-}$ molecular wires.
\end{abstract}

\maketitle
	
\section{Introduction}\label{intro}

\begin{figure}[t]
	\includegraphics[width=0.47\textwidth]{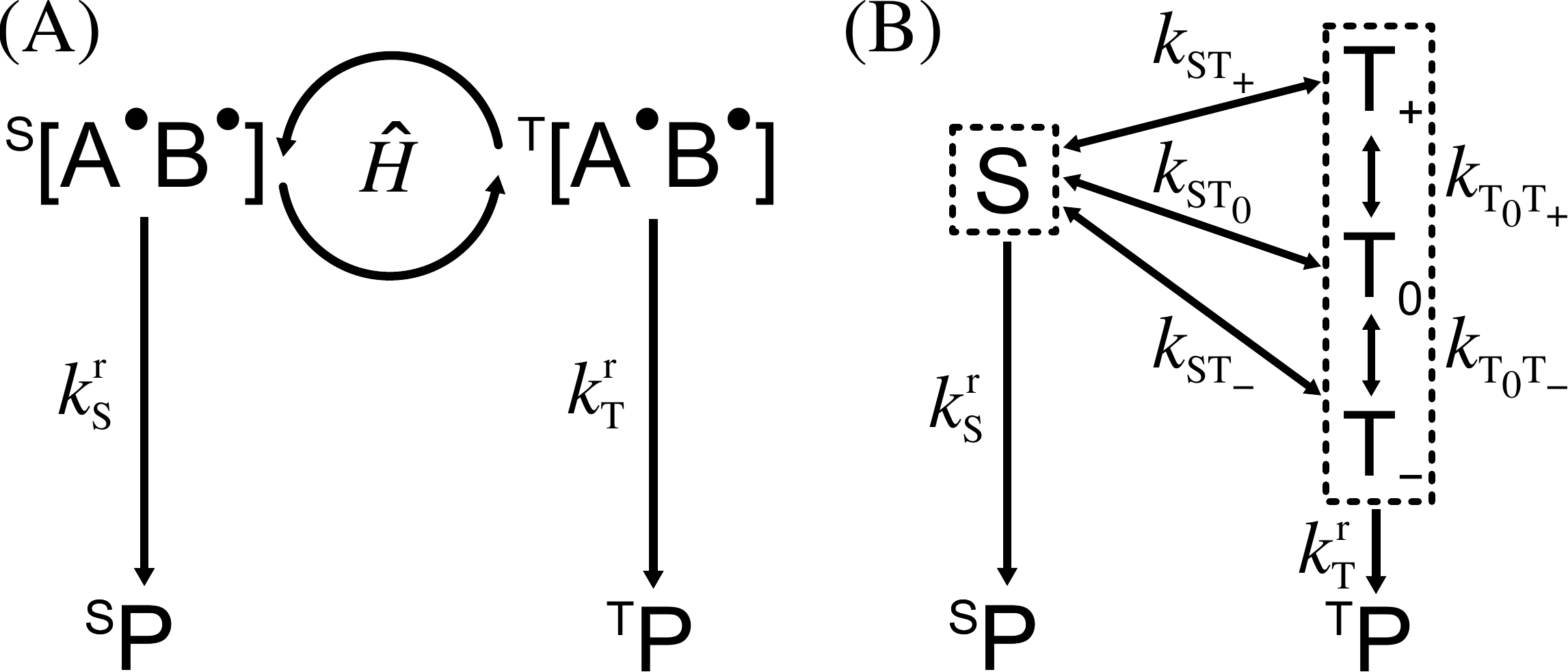}
	\caption{\footnotesize{(A) A schematic picture of the radical pair mechanism with coherent quantum intersystem crossing between singlet and triplet states. (B) The incoherent kinetic scheme used to describe the radical pair mechanism.\label{rp-mech-fig}}}
\end{figure}

Weak magnetic interactions in radical pairs can give rise to extremely large effects on their reactions.\cite{Steiner1989,Rodgers2009} In particular, applied magnetic fields can have significant effects on the extent of intersystem crossing between singlet and triplet states in radical pairs. These effects are usually described using models that include quantum coherences between spin states, based on the Liouville-von Neumann equation for the spin density operator $\op{\rho}(t)$. The action of the effective spin Hamiltonian $\op{H}$ on $\op{\rho}(t)$ gives rise to coherent evolution of spins in the radical pair, as depicted in Fig.~\ref{rp-mech-fig} (A). However, experiments which probe radical pair survival probabilities and the quantum yields of spin state selective recombination reactions are often interpreted using simple incoherent kinetic models for the interconversion of radical pair spin states, as depicted in Fig.~\ref{rp-mech-fig} (B).\cite{Scott2011,Miura2010,Klein2015,Lukzen2017,Steiner2018,Hayashi1984} 

One particular model proposed recently by Steiner \textit{et al.} employs the following functional form for the spin state interconversion rates,\cite{Klein2015,Lukzen2017,Steiner2018}
\begin{align}
k_{nm} = \frac{k_\mathrm{hfc}}{1 + (\epsilon_n-\epsilon_m)^2/\gamma_{\mathrm{hfc}}^2} + \frac{k_\mathrm{rel}}{1 + (\epsilon_n-\epsilon_m)^2/\gamma_{\mathrm{rel}}^2} + k_0, \label{steiner-eqn}
\end{align}
in which $k_\mathrm{hfc}$, $k_\mathrm{rel}$, $k_0$, $\gamma_{\mathrm{hfc}}$ and $\gamma_{\mathrm{rel}}$ are free parameters and $\epsilon_n$ is the energy of \highlight{the coupled electronic spin state $\ket{n}=$ $\ket{\sing}$, $\ket{\trip_{+}}$, $\ket{\trip_0}$, or $\ket{\trip_-}$ in the absence of hyperfine interactions}.\footnote{\highlight{The coupled spin states here have the standard definitions in terms of the uncoupled electron spin states $\ket{\alpha_i}$ and $\ket{\beta_i}$, see for example Ref. \onlinecite{Steiner1989}. $\ket{\sing} = (\ket{\alpha_1\beta_2}-\ket{\beta_1\alpha_2})/\sqrt{2}$, $\ket{\trip_+} = \ket{\alpha_1 \alpha_2}$, $\ket{\trip_0} = (\ket{\alpha_1\beta_2}+\ket{\beta_1\alpha_2})/\sqrt{2}$ and $\ket{\trip_-} = \ket{\beta_1 \beta_2}$.}} Here the first term represents the isotropic hyperfine contribution to the interconversion and the second represents the spin relaxation contribution. This ansatz has been used successfully to interpret the magnetic field effects on radical pair survival probabilities in several sets of experiments.\cite{Klein2015,Lukzen2017,Steiner2018} A similar expression for the hyperfine mediated intersystem crossing rate has previously been arrived at by applying the steady-state approximation to the coherences in a simple two-state model of the radical pair spin states.\cite{Miura2006,Maeda2006,Miura2008,Mojaza2012} 

At a glance, the coherent quantum dynamics approach and the kinetic approach appear to be fundamentally different. But in this paper we shall show how the kinetic model can in fact be derived as an approximation to the exact quantum spin dynamics. In particular, we shall show that expressions for the spin-state interconversion rate constants very similar to those in Eq. \eqref{steiner-eqn} can be obtained straightforwardly from a perturbative approximation to the solution of an appropriate Nakajima-Zwanzig equation.

\section{Theory}
\subsection{Radical pair spin dynamics}

The radical pair state is described by its density operator $\hat{\rho}(t)$, which evolves according to the quantum master equation\cite{Haberkorn1976, Ivanov2010, Fay2018}
\begin{align}
\dv{t}\,\op{\rho}(t) = -\frac{i}{\hbar}\left[\op{H},\op{\rho}(t)\right] - \left\{\frac{\kS}{2}\op{P}_\sing + \frac{\kT}{2}\op{P}_\trip,\op{\rho}(t)\right\} - \pD\op{\rho}(t),\label{haberkorn-qme-eqn}
\end{align}
\highlight{in which $\kS$ and $\kT$ are first-order spin selective recombination rate constants, and $\op{P}_\sing = \dyad{\sing}$, and $\op{P}_\trip = \dyad{\trip_+}+\dyad{\trip_0}+\dyad{\trip_-}$ are projection operators onto the singlet and triplet electronic subspaces.} The first term on the right-hand side of Eq.~\eqref{haberkorn-qme-eqn} describes the coherent spin evolution, the second describes the effect of spin-state selective radical pair recombination reactions and the third describes any additional singlet-triplet (and triplet-triplet) dephasing, 
\begin{align}
\pD\op{\rho}(t) = \sum_{n\neq m} \kD{n}{m} \op{P}_n\,\op{\rho}(t)\op{P}_m,
\end{align}
\highlight{in which $\op{P}_n=\dyad{n}$}. Here $\kD{n}{m}=\kD{m}{n}$ is the additional dephasing rate for the $n,m$ coherence, which arises from fluctuations in the electron spin coupling as a result of nuclear motion\cite{Kattnig2016} or strong diabatic coupling between the radical pair and product states.\cite{Fay2018}

The Hamiltonian $\hat{H}$ in Eq.~(2) can be split into reference part $\op{H}_0$ and a perturbation $\op{V}$. We will take the reference part to include the average Zeeman interaction and scalar electron spin coupling, and the perturbation to include the isotropic nuclear hyperfine couplings and the difference between the Zeeman interactions of the two radicals,
\begin{align}
\op{H}_0 &= \frac{\mu_\mathrm{B}B}{2} (g_1+g_2) (\op{S}_{1z}+\op{S}_{2z}) - 2J \op{\vb{S}}_1\cdot\op{\vb{S}}_2, \\
\op{V} &= \frac{\mu_\mathrm{B}B}{2}(g_1-g_2) (\op{S}_{1z}-\op{S}_{2z}) + \sum_{i=1,2}\sum_{k=1}^{N_i} a_{ik} \op{\vb{I}}_{ik}\cdot\op{\vb{S}}_i.
\end{align}
\highlight{Here $\op{\vb{S}}_i$ is the unitless electron spin operator for radical $i$, $\op{\vb{I}}_{ik}$ is the nuclear spin operator for nucleus $k$ on radical $i$, $a_{ik}$ is the isotropic hyperfine coupling constant for this nucleus, $B$ is the applied magnetic field strength, $\mu_\mathrm{B}$ is the Bohr magneton, $g_i$ is the isotropic g-factor for radical $i$, and $J$ is the scalar coupling constant for the electron spins.\cite{Steiner1989}} 

Using these definitions, we can split the full Liouvillian $\pL$, \highlight{defined by Eq. \eqref{haberkorn-qme-eqn}}, into a reference part $\pL_0$ and a perturbation $\pL_V$. The perturbation is taken to only include the action of $\op{V}$ in Liouville space, $\pL_V = -(i/\hbar)[\op{V},\cdot]$, and the reference is taken to be the remainder of the Liouvillian, $\pL_0 = \pL - \pL_V$, \highlight{including reaction and dephasing terms in Eq. \eqref{haberkorn-qme-eqn}}. \highlight{From the definition of $\ket{n}$ given in Section \ref{intro}, it is straightforward to show that} $\dyad{n}{m}$ is Liouville-space eigenvector of $\pL_0$ with eigenvalue $\lambda_{nm} = -i(\epsilon_n-\epsilon_m) - \gamma_{nm}$. $\epsilon_n$ is the eigenvalue of $\op{H}_0/\hbar$ associated with $\ket{n}$ and $\gamma_{nm}$ is the total decay rate of $\dyad{n}{m}$, $\gamma_{nm}=(k_{n}^{\mathrm{r}}+k_{m}^{\mathrm{r}})/2 + (1-\delta_{nm})\kD{n}{m}$, \highlight{which arises from the reaction and dephasing terms in Eq. \eqref{haberkorn-qme-eqn} (here $k_{n}^{\mathrm{r}}$ is the reaction rate of state $\ket{n}$, either $\kS$ or $\kT$)}. 

The initial radical pair spin density matrix for radical pair reactions can usually be written as a sum of electronic spin state projection operators, 
\begin{align}
\op{\rho}(0) = \frac{1}{Z} \sum_{n}p_n(0) \op{P}_n,
\end{align}
where $p_n(0)$ is the initial probability of finding the radical pair in state $n$ and $Z$ is the dimensionality of the nuclear spin Hilbert space, $Z = \prod_{i=1,2}\prod_{k=1}^{N_i}(2I_{ik}+1)$.

\subsection{The kinetic master equation}

The Nakajima-Zwanzig equation is an exact quantum master equation for the projected density operator $\pP\op{\rho}(t)$,\cite{Nakajima1958,Zwanzig1960}
\begin{align}\label{nz-eqn}
\dv{t}\pP \op{\rho}(t) = \pL_0 \pP \op{\rho}(t) + \int_0^t \dd{\tau} \pK(t-\tau)\pP\op{\rho}(\tau).
\end{align}
The kernel $\pK(t)$ is given by,
\begin{align}
\pK(t) = \pP \pL_V \pQ e^{\pQ \pL t} \pQ \pL_V \pP,
\end{align}
in which $\pQ = 1 -\pP$ and it has been assumed that $\pP \op{\rho}(0) = \op{\rho}(0)$ and $\pP \pL_0 = \pL_0 \pP$. This can be used to obtain a master equation for the populations by defining the projection operator as 
\begin{align}
\pP = \frac{1}{Z}\sum_n\op{P}_n\Tr[\op{P}_n\ \cdot \ ],
\end{align}
and then an exact equation for the populations can be obtained by taking the trace of this projected onto each of the spin states, $p_n(t) = \Tr[\op{P}_n\pP\op{\rho}(t)]$,
\begin{align}
\dv{t} p_n(t) = -k_n^{\mathrm{r}}p_n(t) + \sum_{m}\int_0^t\dd{\tau} \kappa_{nm}(t-\tau)p_m(\tau),
\end{align}
where the rate kernels are given by $\kappa_{nm}(t) = \Tr[\op{P}_n \pL_V \pQ e^{\pQ \pL t} \pQ \pL_V \op{P}_m]$. \highlight{The decay time of $\kappa_{nm}(t)$ dictates the time-scale on which short-time coherent oscillations decay.} If the kernels decay on a time-scale faster than the dynamics of $p_{n}(t)$, we can make the incoherent rate approximation to obtain the Markovian kinetic master equation (KME),\cite{Sparpaglione1988, Fay2018}
\begin{align}
\dv{t} p_n(t) = -k_n^{\mathrm{r}}p_n(t) + \sum_{m} k_{nm}p_m(t),
\end{align}
in which the rate constants $k_{nm}$ are given by $k_{nm} = \int_0^\infty \dd{t} \kappa_{nm}(t)$. 

For time-integrated properties such as the singlet quantum yield, $\Phi_\sing = \kS\int_0^\infty p_\sing(t)\dd{t}$, the KME is exact,\cite{Fay2018} as can be seen by comparing the Laplace transforms of Eqs.~(10) and (11). However, in order to derive explicit expressions for the rate constants $k_{nm}$, we shall now treat $\hat{V}$ as a perturbation. The rate constants can be evaluated to second order in $\op{V}$ by approximating the rate kernels as $\kappa_{nm}(t) \approx \Tr[\op{P}_n \pL_V \pQ e^{\pQ \pL_0 t} \pQ \pL_V \op{P}_m]$. This approximation yields a second order kinetic master equation (KME2) which rigorously gives integrated and long-time properties accurate to second order in the perturbation. From this we can also obtain a criterion of the validity of the kinetic description.

\subsection{Intersystem crossing rate constants}


We can obtain the rate constants $k_{nm}$ by integrating the second order rate kernels. Noting that $e^{\pQ\pL_0 t} = \pP + \pQ e^{\pL_0 t} $, the second order approximation to $\kappa_{nm}(t)$ for $n\neq m$ is
\begin{align}
\kappa_{nm}(t) = \frac{2}{\hbar^2 Z}e^{-\gamma_{nm}t} \cos[(\epsilon_n-\epsilon_m)t]\Tr[\op{P}_n\op{V}\op{P}_m\op{V}],\label{tc2-term-eqn}
\end{align}
and $\kappa_{nn}(t) = -\sum_{m\neq n}\kappa_{mn}(t)$. 
In order to evaluate the kernels, we need the following matrix elements of $\op{V}$, $\mel{n}{\hat{V}}{m}\equiv \op{V}_{nm}=\op{V}_{mn}^\dag$,
\begin{subequations}
\begin{align}
\op{V}_{\sing\trip_\pm} &= \mp\frac{1}{2\sqrt{2}} \left[(\op{h}_{1x}\pm i\op{h}_{1y}) - (\op{h}_{2x}\pm i\op{h}_{2y})\right]\\
\op{V}_{\sing\trip_0} &= \frac{1}{2} \left[\op{h}_{1z} - \op{h}_{2z}\right] + \frac{1}{4}\mu_\mathrm{B}(g_1-g_2) B \\
\op{V}_{\trip_0\trip_{\pm}} &= \frac{1}{2\sqrt{2}} \left[(\op{h}_{1x}\pm i\op{h}_{1y}) + (\op{h}_{2x}\pm i\op{h}_{2y})\right] \\
\op{V}_{\trip_{\pm}\trip_{\mp}} &= 0,
\end{align}
\end{subequations}
in which $\op{h}_{i\alpha} = \sum_{k=1}^{N_i}a_{ik}\op{I}_{ik\alpha}$. Taking the trace of products these as in Eq.~\eqref{tc2-term-eqn}, it is clear that the only non-vanishing terms are those proportional to an $\op{I}_{ik\alpha}^2$ or an identity operator. The trace of $\op{I}_{ik\alpha}^2$ is $\Tr_\mathrm{nuc}[\op{I}_{ik\alpha}^2] = \frac{1}{3}I_{ik}(I_{ik}+1)Z$, which can be used to evaluate all of the terms appearing in the master equation.

The kinetic master equation rate constants satisfy $k_{nm} = k_{mn}$ and they can be split into the sum of a hyperfine contribution $k_{nm}^{(\mathrm{hfc})}$ and a $\Delta g$ contribution $k_{nm}^{(\Delta \mathrm{g})}$. It is clear that $k_{\trip_{\pm}\trip_{\mp}}^{(\mathrm{hfc})}=0$, and that all $\Delta g$ rate constants are zero other than $k_{\sing\trip_0}^{(\mathrm{\Delta g})}=k_{\trip_0\sing}^{(\mathrm{\Delta g})} $. The non-zero rate constants are
\begin{align}
k_{nm}^{(\mathrm{hfc})} &=\frac{\omega_{1,\mathrm{hyp}}^2+\omega_{2,\mathrm{hyp}}^2}{6}\frac{\gamma_{nm}}{\gamma_{nm}^2 + (\epsilon_n-\epsilon_m)^2},\label{hfc-rate-eqn} \\
k_{\sing\trip_0}^{(\mathrm{\Delta g})} &= \frac{1}{2}\left(\frac{\mu_\mathrm{B}(g_1-g_2) B}{2\hbar}\right)^2\frac{\gamma_{\sing\trip_0}}{\gamma_{\sing\trip_0}^2 + (\epsilon_\sing-\epsilon_{\trip_0})^2},\label{dg-date-eqn}
\end{align}
where $\omega_{i,\mathrm{hyp}}^2 = \sum_{k=1}^{N_i} a_{ik}^2 I_{ik}(I_{ik}+1)/\hbar^2$. 

The generalisation of the master equation to include electron spin relaxation arising from rotational diffusion is straightforward. Here we shall simply state the additional contributions to the singlet-triplet interconversion rates for a radical pair undergoing isotropic rotational diffusion, and leave the details of the derivation to the Supplementary Material. 
The final expression for the relaxation-induced spin-state interconversion rates is
\begin{align}
k_{nm}^{(\mathrm{hf-aniso})}\! &=\! \frac{|\omega_{1,\mathrm{hyp}}^{(2)}|^2+|\omega_{2,\mathrm{hyp}}^{(2)}|^2}{18} \frac{(\gamma_{nm}+{1/\tau_\mathrm{R}})}{(\gamma_{nm}\!+\!{1/\tau_\mathrm{R}})^2\!+\!(\epsilon_n\!-\!\epsilon_{m})^2},\label{hf-aniso-rate-eqns}
\end{align}
where $\tau_{\rm R}$ is the isotropic rotational correlation time,\cite{Lau2010} $|\omega_{i,\mathrm{hyp}}^{(2)}|^2 = \sum_{k=1}^{N_i}\sum_{m=-2}^2|A_{ik,m}^{(2)}|^2I_{ik}(I_{ik}+1)/\hbar^2$, and $A_{ik,m}^{(2)}$ is the $m^\text{th}$ rank 2 spherical tensor component of the hyperfine coupling for nuclear spin $k$ on radical $i$.\cite{Nicholas2010} Analogous expressions for relaxation induced by rotational modulation of g-tensor anisotropy are given in the Supplementary Material, including the effect of anisotropic rotational diffusion. 

Eqs.~\eqref{hfc-rate-eqn} to~\eqref{hf-aniso-rate-eqns} are clearly very closely related to the ansatz proposed by Steiner \textit{et al.}\cite{Steiner2018} [Eq.~\eqref{steiner-eqn}]. However, they have been derived here directly from the quantum mechanical description of the radical pair spin dynamics, and they do not involve any free parameters. One significant difference between our equations and Eq.~(1) is that our width parameters $\gamma_{nm}$ depend explicitly on the spin states $n$ and $m$ that are interconverting. 

\highlight{From the theory outlined above, we can find criteria for the validty of the Markovian and perturbative approximations. The second order perturbative appoximation will be valid when the time-scales of the unperturbed dynamics are shorter than that of the perturbed dynamics, i.e. for the isotropic hyperfine interactions when $(\omega_{1,\mathrm{hyp}}^2+\omega_{2,\mathrm{hyp}}^2)/{6}\ll\gamma_{nm}^2 + (\epsilon_n-\epsilon_m)^2$, for the $\Delta$g mechanism when $({\mu_\mathrm{B}(g_1-g_2) B}/{\hbar})^2/8\ll\gamma_{\sing\trip_0}^2 + (\epsilon_\sing-\epsilon_{\trip_0})^2$, and for the anisotropic hyperfine interactions when $({|\omega_{1,\mathrm{hyp}}^{(2)}|^2+|\omega_{2,\mathrm{hyp}}^{(2)}|^2})/{18}\ll(\gamma_{nm}+1/\tau_{\mathrm{R}})^2 + (\epsilon_n-\epsilon_m)^2$, where $n\neq m$. The Markovian approximation will be valid when the decay time of the kernels is shorter than the time-scale of the population dynamics. This means that the Markovian approximation will be valid for the isotropic interactions when $k_{nm}^{(\mathrm{hfc/\Delta g})}<\gamma_{nm}$ and for the anisotropic interactions when $k_{nm}^{(\mathrm{hf-aniso})}<\gamma_{nm}+1/\tau_{\mathrm{R}}$. These criteria for Markovianity are the same as the criteria for the validity of second order perturbation theory. Higher order truncations of the kernel in Eq. \eqref{nz-eqn} and approximate resummations of these higher order terms could in principle be used to obtain master equations valid beyond the perturbative limit,\cite{Sparpaglione1988,Fay2018} however the resulting rate constants would have a significantly more complex functional form than that proposed by Steiner \textit{et al.} [Eq. \eqref{steiner-eqn}].}

\section{Example systems}

\begin{figure}[t]
	\includegraphics[width=0.49\textwidth]{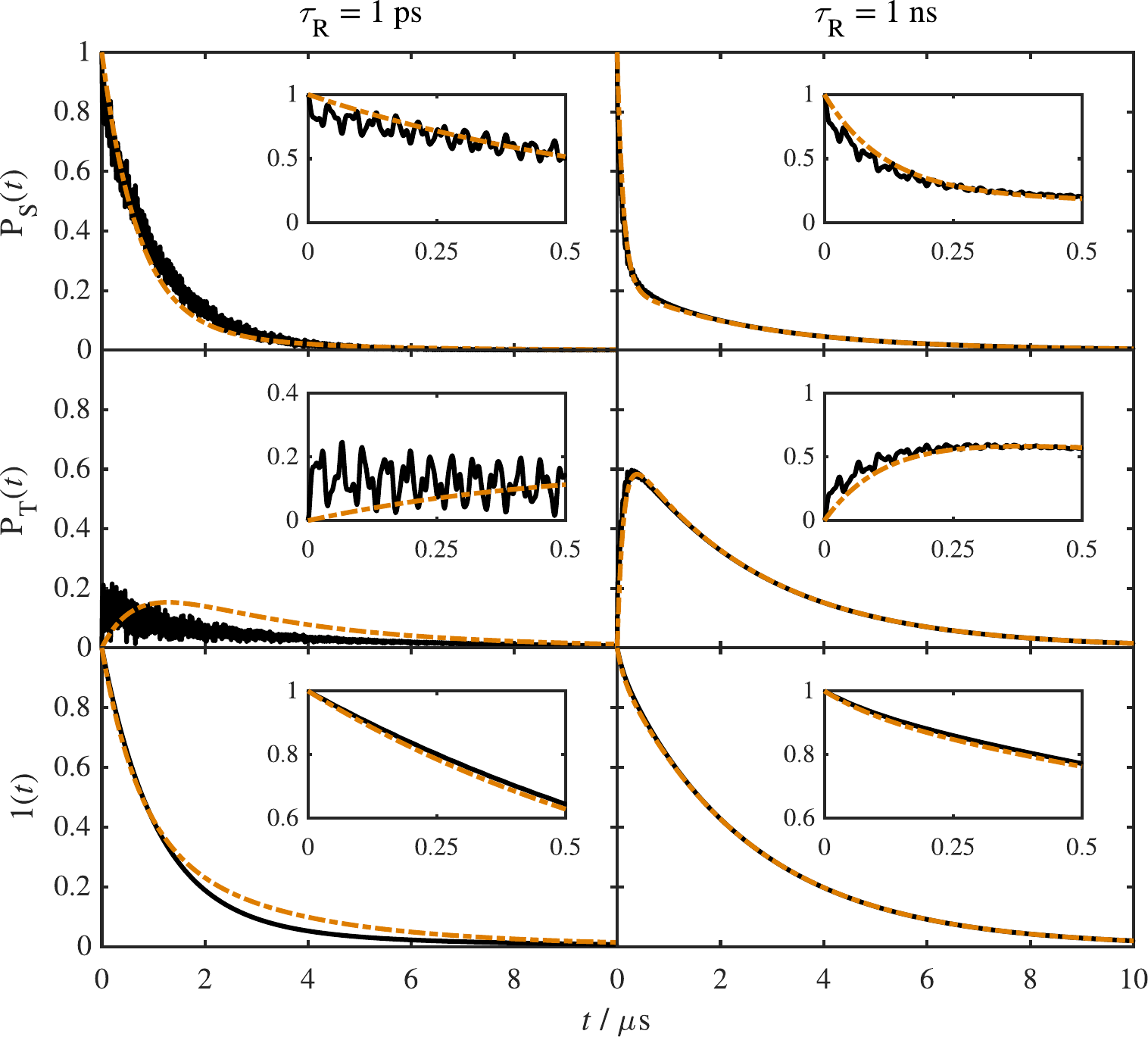}
	\caption{\footnotesize{Singlet, triplet and total survival probabilities for a one proton radical pair with isotropic rotational diffusion causing relaxation. Black solid lines are the exact probabilities obtained by solving the Stochastic Liouville equation and the orange dot-dash lines are the KME2 results. The insets show the probabilities for short times (up to $0.5\ \mu\text{s}$). Both radicals have isotropic g-tensors with $g_i = g_\el$ and are in an external field of strength $B = 1\text{ mT}$. The scalar coupling is $J = -0.75\text{ mT}$ and the proton has a diagonal hyperfine coupling tensor with components $A_{xx}=A_{yy}=0.5\text{ mT}$ and $A_{zz}=2\text{ mT}$. The radical pair reacts asymmetrically with $k_\sing^\mathrm{r} = 1\ \mu\text{s}^{-1}$ and $k_\trip^\mathrm{r} = 0.2\ \mu\text{s}^{-1}$.} \label{one-proton-fig}}
\end{figure}
\begin{figure}
	\includegraphics[width=0.47\textwidth]{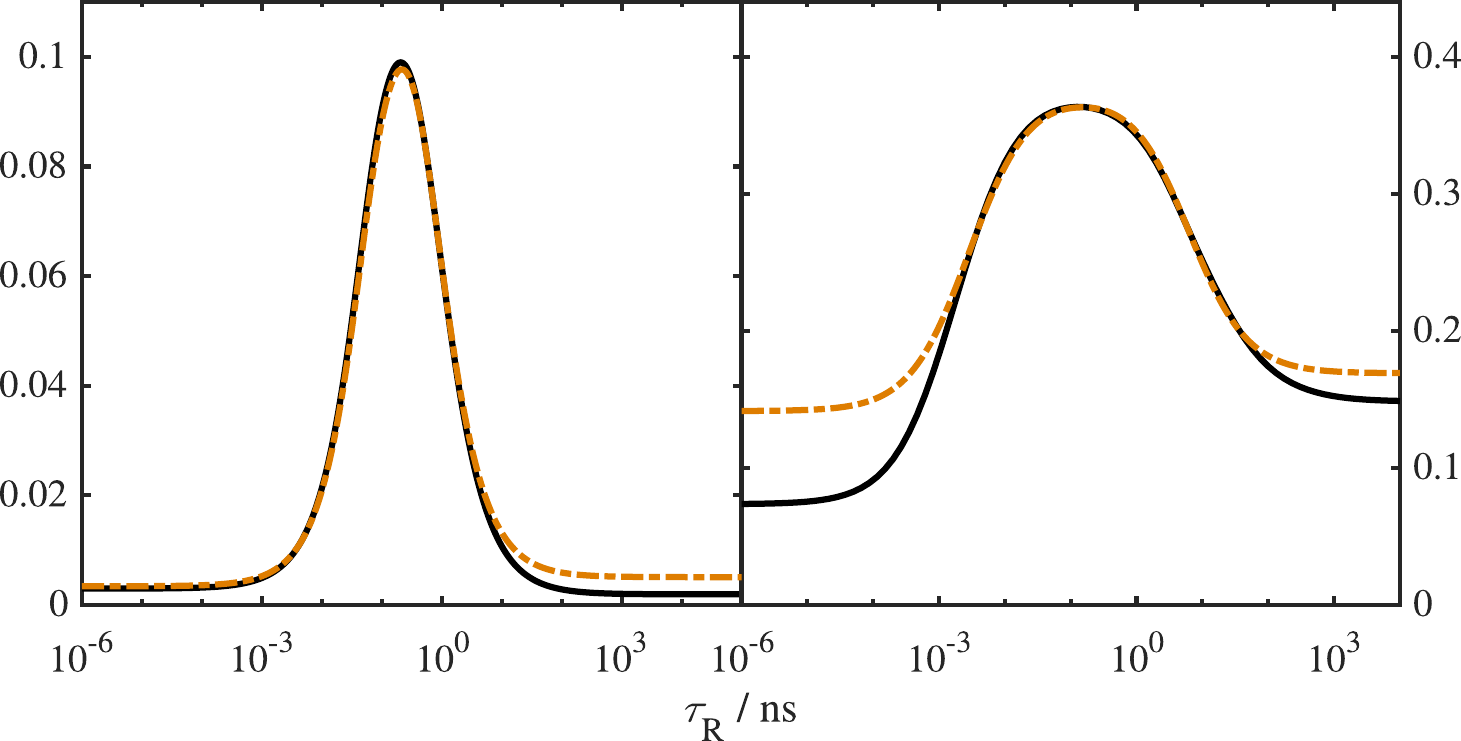}
	\caption{\footnotesize{Triplet yield as a function of rotational correlation time for a one proton radical pair. The simulations in the right panel used same parameters for the radical pair as in Fig.~\ref{one-proton-fig}. In the left panel the hyperfine coupling tensor has been reduced by a factor of 10 to $A_{xx}\!=\!A_{yy}\!=\!0.05\text{ mT}$ and $A_{zz}\!=\!0.2\text{ mT}$. \label{phi-T-fig}}}
\end{figure}
\begin{figure}[t]
	\includegraphics[width=0.5\textwidth]{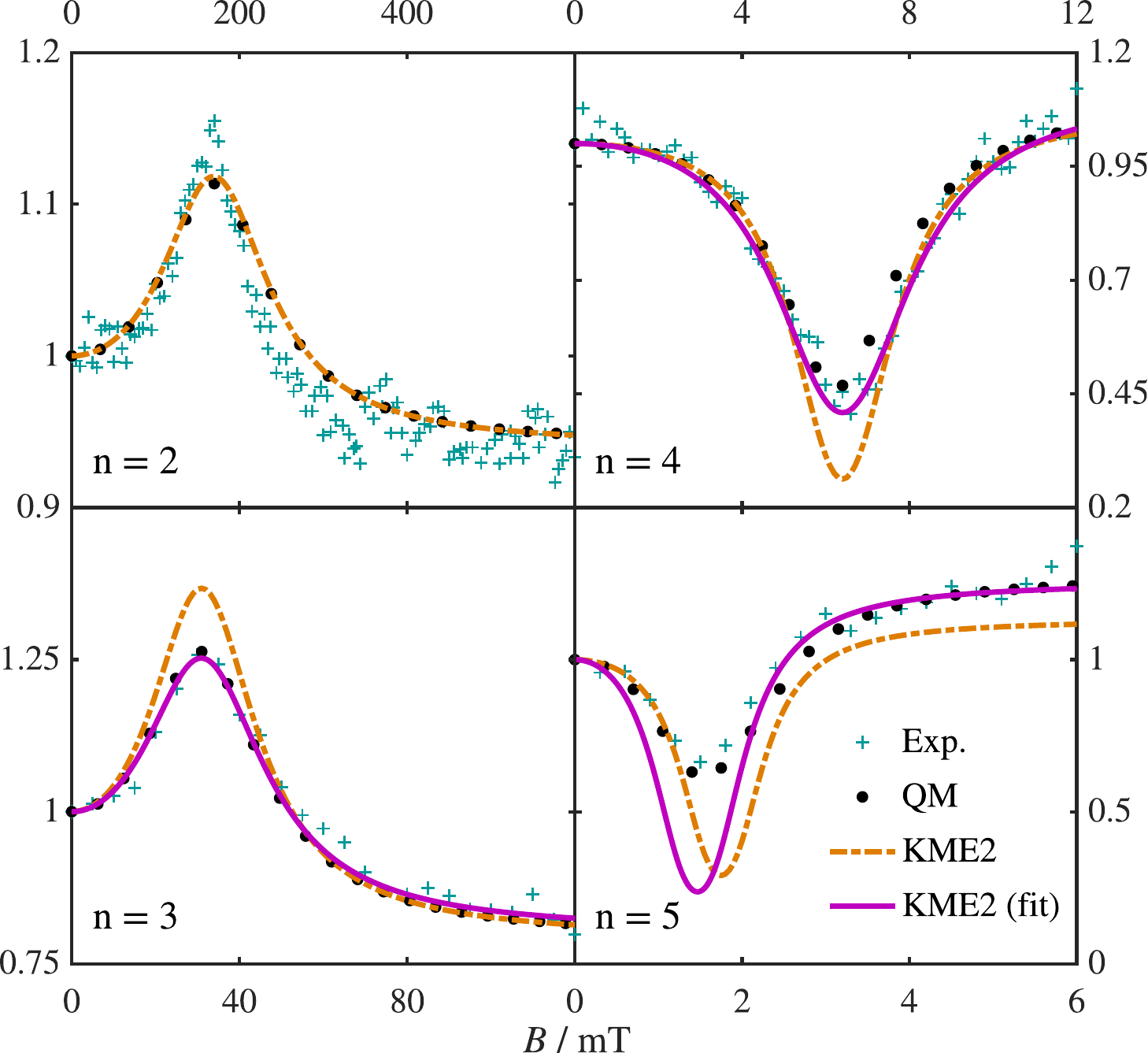}
	\caption{\footnotesize{Simulated magnetic field effects for a series of $\text{PTZ}^{\bullet+}\text{--Ph}_\text{n}\text{--PDI}^{\bullet-}$ molecular wires. The n=2 and n=3 panels show relative triplet yields and the n=4 and n=5 panels show relative survival probabilities of the radical pair at $t=55\text{ ns}$. The experimental data to which the quantum simulations were originally fitted are also include for comparison.\cite{Weiss2004} The QM and KME2 simulations used identical parameters, including the same background correction for the n=2 and n=3 data (see Ref. \onlinecite{Fay2017} for more details). The KME2 (fit) results were produced by refitting the model parameters.} \label{ptz-pdi-fig}}
\end{figure}

In order to evaluate the accuracy of the perturbative master equation, and in particular to demonstrate where the approximations we have made in deriving it are not applicable, we will now present calculations for two example systems for which exact quantum mechanical simulations can be performed for comparison, using either the stochastic Liouville equation\cite{Lau2010} or spin coherent state sampling.\cite{Lewis2016,Fay2017}

\subsection{A single proton radical pair}

As a first example, which includes the effects of electron-spin relaxation, we have simulated the population dynamics for a radical pair undergoing isotropic rotational diffusion, with one radical coupled anisotropically to a single proton. Fig.~\ref{one-proton-fig} shows the singlet, triplet and total survival probabilities for this radical pair with rotational correlation times of 1 ps and 1 ns. In this example $a = B = (-4/3)J = 1.76\times10^2 k_\sing^\mathrm{r}/\gamma_\el=8.8\times10^2 k_\trip^\mathrm{r}/\gamma_\el$, so this is in a regime where the hyperfine coupling is of comparable strength to the electron spin coupling. Furthermore, in the case of only one hyperfine coupled proton, coherence effects will be highly pronounced.\cite{Manolopoulos2013,Lewis2014} As the rotational correlation time decreases from 1 ns to 1 ps, it can be seen that the master equation becomes less accurate. When $\tau_\mathrm{R}$ is small, the second order kinetic master equation obviously fails to capture the coherent oscillations between the singlet and triplet states at short times. However, for longer times, and when relaxation plays a significant role, as in the $\tau_\mathrm{R}=1\text{ ns}$ case, KME2 is very accurate. In this case it can be seen that coherent oscillations decay after $t\approx 0.5\ \mu\text{s}\approx 1/\gamma_{\sing\trip_m}$, which is the decay time of the singlet-triplet rate kernels.

In Fig.~\ref{phi-T-fig} we examine the accuracy of KME2 for the triplet quantum yield of this one proton radical pair, as a function of the rotational correlation time. When the hyperfine coupling is weak, $A_{xx}=A_{yy}=0.05\text{ mT}$ and $A_{zz}=0.2\text{ mT}$ (Fig.~\ref{phi-T-fig}, left panel), KME2 is very accurate up to long correlation times, $\tau_\mathrm{R}> 10\text{ ns}$. For stronger hyperfine coupling,  $A_{xx}=A_{yy}=0.5\text{ mT}$ and $A_{zz}=2\text{ mT}$ (Fig.~\ref{phi-T-fig}, right panel), the perturbative approximation to the hyperfine coupling breaks down, and when relaxation does not contribute to the radical pair intersystem crossing ($\tau_\mathrm{R} < 1 \text{ ps}$) the KME2 results no longer agree quantitatively with the exact results. \highlight{This breakdown arises because the perturbation strength, $\omega_{1,\mathrm{hyp}}/\sqrt{6} = a / (2\sqrt{2})\approx 0.35 \text{ mT}$, is approximately the same as the smallest unperturbed frequency, $\min_{n\neq m}[{\gamma_{nm}^2+(\epsilon_n-\epsilon_{m})^2}]^{1/2}\approx 0.5 \text{ mT}$, consistent with the above discussion.} The second order master equation nevertheless remains accurate when relaxation dominates (for 1 ps $<\tau_{\rm R}<$ 100 ns), in spite of the strong hyperfine coupling.

\subsection{Para-phenylene molecular wires}

As a second example, we have simulated magnetic field effects on the recombination reactions of a homologous series of charge-separated $\text{PTZ}^{\bullet+}\text{--Ph}_\text{n}\text{--PDI}^{\bullet-}$ molecular wires.\cite{Weiss2004} We have previously studied this series using exact quantum mechanical simulations, as described in Ref.~\onlinecite{Fay2017}, and here we use the same model and parameters for the radical pair spin dynamics. In this model no relaxation contributions are present, so one would expect the Markovian and weak-coupling approximations to be less accurate. From n=2 to n=5, the total scalar electron spin coupling decreases from $|2J|=170\text{ mT}$ to $|2J|=1.75\text{ mT}$, \highlight{which is comparable to the perturbation strength, $[(\omega_{1,\mathrm{hyp}}^2+\omega_{2,\mathrm{hyp}}^2)/6]^{1/2}\approx 0.41\text{ mT}$}, and so treating the hyperfine interactions to lowest order in perturbation theory breaks down along the series, as can be seen from the results in Fig.~\ref{ptz-pdi-fig}. The largest deviations occur when there is a near degeneracy between the $\sing$ and $\trip_{-}$ states at $B=2J$, where the perturbative approximation is least valid, and in the n=5 case where the low field effect\cite{Lewisa} contributes significantly to the magnetic field effect on the radical pair survival probability.\cite{Fay2017} Despite this the errors in the KME2 results are still relatively small for the n=2--4 radical pairs. 

For the n=3--5 molecular wires we have used the kinetic master equation to fit the experimental data using the same free parameters as in the quantum simulations described in Ref.~\onlinecite{Fay2017}. These fits are also shown in Fig.~\ref{ptz-pdi-fig}. In the n=3 and n=4 cases the data can be fit well with KME2, with parameters differing from the QM parameters by $<20\%$ for n=3 and $<50\%$ for n=4. However, for the n=5 molecule, we were unable to find $k_\sing^\mathrm{r},\ k_\trip^\mathrm{r}\text{ and }J$ parameters for which the KME2 approximation gave a good fit to the experimental data. This is because of the importance of the low magnetic field effect in this case, and the value small of $|2J|$. Under these circumstances the KME2 is clearly inadequate, and the only reliable way we know of to fit the experimental data is to resort to a coherent quantum mechanical calculation of the type described in Ref.~\onlinecite{Fay2017}.

\section{Concluding Remarks}

In this paper, we have shown how the incoherent kinetic description of radical pair intersystem crossing can be derived  from quantum dynamics. A perturbative approximation to the nuclear hyperfine coupling in the exact Nakajima-Zwanzig equation leads to a second order kinetic master equation for the electronic spin state populations. It is seen that the finite lifetime of the radical pair spin states, as well as any additional dephasing, drives the transition to incoherent kinetic behaviour. The KME2 is accurate in the long-time limit and exact for time integrated properties exactly to lowest order in the hyperfine interactions and in the difference between the radical g-tensors. 
Tests on model systems have shown that the simple kinetic equations are remarkably accurate when the singlet-triplet coherence time is short, or when relaxation processes dominate, and when the hyperfine interaction is relatively weak compared to other spin interactions. However, the second order kinetic description obviously has some shortcomings. For example, it fails to capture the decrease in the $\sing$ to $\trip_0$ interconversion rate at low applied magnetic field strengths,\cite{Lewisa} as demonstrated by the failure of KME2 to quantitatively capture the magnetic field effect on the survival probability of the $\text{PTZ}^{\bullet+}\text{--Ph}_\text{5}\text{--PDI}^{\bullet-}$ radical pair. There are however many situations in which the approximation works well (see Figs.~2-4), and so we expect that the theory developed here will prove useful in the interpretation of many future experiments on radical pair reactions.

\section*{Supplementary Material}

In the Supplementary Material we outline the derivation of the rotational diffusion contributions to the spin-state interconversion rates from the Nakajima-Zwanzig equation, including the effects of anisotropic rotational diffusion and g-tensor anisotropy, and give all of the parameters used in the $\text{PTZ}^{\bullet+}\text{--Ph}_\text{n}\text{--PDI}^{\bullet-}$ spin dynamics simulations.

\begin{acknowledgements}
	
	Thomas Fay is supported by a Clarendon Scholarship from Oxford University, an E.A. Haigh Scholarship from Corpus Christi College, Oxford, and by the EPRSC Centre for Doctoral Training in Theory and Modelling in the Chemical Sciences, EPSRC Grant No. EP/L015722/1. 
	
\end{acknowledgements}

\bibliography{kinetic-equations-v2}

\end{document}


\title{Supplementary Material to ``Radical pair intersystem crossing: quantum dynamics or incoherent kinetics?''}
\author{Thomas P. Fay}
\email{thomas.fay@chem.ox.ac.uk}
\affiliation{Department of Chemistry, University of Oxford, Physical and Theoretical Chemistry Laboratory, South Parks Road, Oxford, OX1 3QZ, UK}
\author{David E. Manolopoulos}
\affiliation{Department of Chemistry, University of Oxford, Physical and Theoretical Chemistry Laboratory, South Parks Road, Oxford, OX1 3QZ, UK}

\begin{abstract}
	In this supplementary material we outline the derivation of the relaxation contributions to incoherent spin-state interconversion rates for fully anisotropic rotational diffusion of the radical pair. We also give all of the parameters used in the simulations of the $\text{PTZ}^{\bullet+}\text{--Ph}_\text{n}\text{--PDI}^{\bullet-}$ radical pair reactions.
\end{abstract}

\maketitle

\section{The kinetic master equation with rotational diffusion}

In the following we assume the radical pair is rigid, and the orientation between the radicals is fixed. The case of two independently rotating radicals is straightforward. Our starting point for deriving the kinetic equations for radical pair intersystem crossing including the effects of rotational diffusion is the Stochastic Liouville equation for the radical pair spin density operator, as a function of time $t$ and orientation of the molecule $\Omega$\cite{Lau2010}
\begin{align}
\dv{t}\op{\rho}(t,\Omega) = -\frac{i}{\hbar}\left[\op{H}(\Omega),\op{\rho}(t,\Omega)\right] - \left\{\frac{\kS}{2}\op{P}_\sing + \frac{\kT}{2}\op{P}_\trip,\op{\rho}(t,\Omega)\right\} - \sum_{n\neq m} \kD{n}{m} \op{P}_n\op{\rho}(t,\Omega)\op{P}_m - \mathsf{D}\op{\rho}(t,\Omega).
\end{align}
where $\mathsf{D}$ is the rotational diffusion operator
\begin{align}
\mathsf{D} = D_X \mathsf{L}_X^2 + D_Y \mathsf{L}_Y^2 + D_Z \mathsf{L}_Z^2
\end{align}
where $D_A$ is the rotational diffusion constant and $\mathsf{L}_A$ is the rotational angular momentum operator about the body-fixed axis $A$. Including the rotational diffusion, we take the reference part of the Hamiltonian $\op{H}_0$ to be defined as before and now we also include the anisotropic spin couplings in the perturbation $\op{V}(\Omega)$, which also defines $\pL_V$,
\begin{align}
\op{V}(\Omega) = \frac{\mu_\mathrm{B}B}{2}(g_1-g_2) (\op{S}_{1z}-\op{S}_{2z}) + \sum_{i=1,2}\sum_{k=1}^{N_i} a_{ik} \op{\vb{I}}_{ik}\cdot\op{\vb{S}}_i + \sum_{m=-2}^2\sum_{m'=-2}^2 \mathfrak{D}_{m',m}^{(2)}(\Omega)\op{Q}_{m,m'}^{(2)}.
\end{align}
$\mathfrak{D}_{m',m}^{(2)}(\Omega)$ is a Wigner D-matrix element and the operators $\op{Q}_{m,m'}^{(2)}$ contain the anisotropic spin couplings,\cite{Nicholas2010}
\begin{align}
\op{Q}_{m,m'}^{(2)} = \sum_{i=1,2}\mu_\mathrm{B} g_{i,m}^{(2)}T_{m'}^{(2)}(\op{\vb{S}}_i,\vb{B})+\sum_{i=1,2}\sum_{k=1}^{N_i}A_{i,k,m}^{(2)}T_{m'}^{(2)}(\op{\vb{S}}_i,\op{\vb{I}}_{i,k}).
\end{align}
$T_{m}^{(2)}(\op{\vb{u}},\op{\vb{v}})$ is component $m$ of a rank 2 spherical tensor for two vector operators, $A_{i,k,m}^{(2)}$ is the $m$th spherical tensor component $m$ of the rank 2 hyperfine coupling tensor for nuclear spin $k$ on radical $i$ and similarly $g_{i,m}^{(2)}$ is a spherical tensor component of the rank 2 g-tensor for radical $i$. The spin density operator will initially be in a state of the form,
\begin{align}
\op{\rho}(0,\Omega) = \op{\rho}(0) p_0(\Omega).
\end{align}
In solution, the initial distribution of the orientations is isotropic
\begin{align}
p_0(\Omega) = \frac{1}{8\pi^2}.
\end{align}
As before we define a Liouville space projection operator which we can use to obtain the kinetic master equation,
\begin{align}
\pP =  \frac{1}{Z}\sum_n p_0(\Omega)\,\op{P}_n \int\dd{\Omega}\,\Tr[\op{P}_n\ \cdot\ ].
\end{align}
In order to evaluate the second order rate kernels, we note that the $\pL_0$ propagator can be split into a spin part and a diffusion part 
\begin{align}
e^{\pL_0  t} = e^{\pL_\mathrm{s}t}e^{-\mathsf{D}t}.
\end{align}
We will also therefore need the rank 2 Wigner D-matrix correlation functions,
\begin{align}
\ev{\mathfrak{D}_{a,b}^{(2)}(\Omega)^*\mathfrak{D}_{a',b'}^{(2)}(\Omega,t) } &= \int\dd{\Omega}\mathfrak{D}_{a,b}^{(2)}(\Omega)^*e^{-\mathsf{D}t}\mathfrak{D}_{a',b'}^{(2)}(\Omega)p_0(\Omega)\nonumber \\
&= \frac{\delta_{a,a'}}{5}\sum_{k=1}^5 h_{k,b}h_{k,b'}^* e^{-  t/\tau_{\mathrm{R},k}}\\
&= \frac{\delta_{a,a'}}{5} c_{b,b'}(t)\nonumber 
\end{align}
where the values of $1/\tau_{\mathrm{R},k}$ are given in Table \ref{lambdak-tab} and the values of $h_{k,m}$ are given in Table \ref{hkm-tab}. We also note that $\ev{\mathfrak{D}_{n,m}^{(2)}(\Omega,t)}=0$. With this we see the terms in the second order kernel that mix isotropic and anisotropic hyperfine couplings vanish, and the anisotropic coupling terms are
\begin{align*}
\kappa_{nm}^{(\mathrm{hf-aniso})}(t) &= \frac{2}{\hbar^2}e^{-\gamma_{nm}t}\cos((\epsilon_n-\epsilon_m)t) \sum_{a,b,a',b'}\ev{\mathfrak{D}_{a,b}^{(2)}(\Omega)^*\mathfrak{D}_{a',b'}^{(2)}(\Omega,t) }\Tr[\op{P}_n \op{Q}_{b,a}^\dag \op{P}_m \op{Q}_{b',a'}] \\
&= \frac{2}{\hbar^2}e^{-\gamma_{nm}t}\cos((\epsilon_n-\epsilon_m)t) \sum_{a,b,b'}\ev{\mathfrak{D}_{a,b}^{(2)}(\Omega)^*\mathfrak{D}_{a,b'}^{(2)}(\Omega,t) }\sum_{i=1,2}\sum_{k=1}^{N_i}{A_{i,k,b}^{(2)*}}A_{i,k,b'}^{(2)}\Tr[\op{P}_nT_{a}^{(2)}(\op{\vb{S}}_i,\op{\vb{I}}_{i,k})^\dag\op{P}_mT_{a}^{(2)}(\op{\vb{S}}_i,\op{\vb{I}}_{i,k})^\dag] \\
&= \frac{2}{\hbar^2}e^{-\gamma_{nm}t}\cos((\epsilon_n-\epsilon_m)t) \sum_{b,b'}\frac{c_{b,b'}(t)}{5} \sum_{i=1,2}\sum_{k=1}^{N_i}{A_{i,k,b}^{(2)*}}A_{i,k,b'}^{(2)}\sum_{a}\Tr[\op{P}_nT_{a}^{(2)}(\op{\vb{S}}_i,\op{\vb{I}}_{i,k})^\dag\op{P}_mT_{a}^{(2)}(\op{\vb{S}}_i,\op{\vb{I}}_{i,k})^\dag]  \\
&= \frac{2}{\hbar^2}e^{-\gamma_{nm}t}\cos((\epsilon_n-\epsilon_m)t) \sum_{b,b'}\frac{c_{b,b'}(t)}{5} \sum_{i=1,2}\sum_{k=1}^{N_i}{A_{i,k,b}^{(2)*}}A_{i,k,b'}^{(2)}\frac{5}{36} I_{i,k}(I_{i,k}+1) \\
&= \frac{1}{18\hbar^2}e^{-\gamma_{nm}t}\cos((\epsilon_n-\epsilon_m)t) \sum_{b,b'}{c_{b,b'}(t)} \sum_{i=1,2}\sum_{k=1}^{N_i}{A_{i,k,b}^{(2)*}}A_{i,k,b'}^{(2)} I_{i,k}(I_{i,k}+1).
\end{align*}
Integrating this gives, for $n\neq m$ 
\begin{align}
k_{nm}^{(\mathrm{hf-aniso})} = \int_0^{\infty} \kappa_{nm}^{(\mathrm{hf-aniso})}(t)\,\dd t = \frac{1}{18}\sum_{j} \frac{\gamma_{nm}+1/\tau_{\mathrm{R},j}}{(\gamma_{nm}+1/\tau_{\mathrm{R},j})^2+(\epsilon_n-\epsilon_m)^2} \sum_{b,b'}h_{j,b}h_{j,b'} \sum_{i=1,2}\sum_{k=1}^{N_i}\frac{{A_{i,k,b}^{(2)*}}A_{i,k,b'}^{(2)} I_{i,k}(I_{i,k}+1)}{\hbar^2},
\end{align}
apart from $k_{\trip_\pm\trip_\mp}^{(\mathrm{hf-aniso})}$, which are zero. 
\begin{table}[t]
	\begin{tabular}{l|c|c|c|c|c}
		$k$ & 1 & 2 & 3 & 4 & 5 \\
		\hline 
		$1/\tau_{\mathrm{R},k}$ & $4D_X+D_Y+D_Z$ & $D_X+4D_Y+D_Z$ & $D_X+D_Y+4D_Z$ & $6\bar{D} - 2 \Delta_D$ & $6\bar{D} + 2 \Delta_D$
	\end{tabular}
\caption{Table of $1/\tau_{\mathrm{R},k}$ values where $\bar{D} = (D_X+D_Y+D_Z)/3$ and $\Delta_D = \sqrt{D_X^2 + D_Y^2 + D_Z^2 - D_X D_Y - D_Y D_Z - D_Z D_X }$. \label{lambdak-tab}}
\end{table}
\begin{table}[t]
	\begin{tabular}{l|c|c|c|c|c}
		{$k$}\textbackslash{$b$}& -2 & -1 & 0 & 1 & 2 \\
		\hline 
		1 & 0 & $1/\sqrt{2}$ & 0 & $1/\sqrt{2}$ & 0 \\
		\hline
		2 & 0 & $-1/\sqrt{2}$ & 0 & $1/\sqrt{2}$ & 0 \\
		\hline
		3 &  $-1/\sqrt{2}$ & 0 & 0 & 0 & $1/\sqrt{2}$ \\
		\hline
		4 & $1/\sqrt{2+\Lambda_-^2}$ & 0 & $\Lambda_-/\sqrt{2+\Lambda_-^2}$ & 0 & $1/\sqrt{2+\Lambda_-^2}$\\
		\hline
		5 & $1/\sqrt{2+\Lambda_+^2}$ & 0 & $\Lambda_+/\sqrt{2+\Lambda_+^2}$ & 0 & $1/\sqrt{2+\Lambda_+^2}$
	\end{tabular}
	\caption{ Table of $h_{k,b}$ values where $\Lambda_{\pm} = \sqrt{2/3}(D_{X}+D_Y - 2D_Z \pm 2\Delta_D)/(D_X-D_Y)$.\label{hkm-tab}}
\end{table}
In the case of symmetric top rotational diffusion, where $D_X = D_Y = D_\perp$ and $D_Z = D_\parallel$, this simplifies to
\begin{align}
k_{nm}^{(\mathrm{hf-aniso})} = \frac{1}{18}\sum_{b} \frac{\gamma_{nm}+1/\tau_{\mathrm{R},b}^{(\mathrm{symm})}}{(\gamma_{nm}+1/\tau_{\mathrm{R},b}^{(\mathrm{symm})})^2+(\epsilon_n-\epsilon_m)^2} \sum_{i=1,2}\sum_{k=1}^{N_i}\frac{|{A_{i,k,b}^{(2)}}|^2 I_{i,k}(I_{i,k}+1)}{\hbar^2}
\end{align}
where $1/\tau_{\mathrm{R},b}^{(\mathrm{symm})} = 6 D_\perp + b^2 (D_\parallel - D_\perp)$. This clearly reduces to the expression given in the main text [Eq.~(16)] when $D_X = D_Y = D_Z$. Similar expressions can be obtained for the g-tensor anisotropy,
\begin{align}
k_{\sing\trip_0}^{(\mathrm{g-aniso})} &= k_{\trip_0\sing}^{(\mathrm{g-aniso})} = \frac{\mu_\mathrm{B}^2 B^2}{15\hbar^2}\sum_{j} \frac{\gamma_{\sing\trip_0}+1/\tau_{\mathrm{R},j}}{(\gamma_{\sing\trip_0}+1/\tau_{\mathrm{R},j})^2+(\epsilon_\sing-\epsilon_{\trip_0})^2} \sum_{b,b'}h_{j,b}h_{j,b'} {\Delta g_{b}^{(2)*}\Delta g_{b'}^{(2)} } \\
k_{\sing\trip_\pm}^{(\mathrm{g-aniso})} &=k_{\trip_\pm \sing}^{(\mathrm{g-aniso})}  = \frac{\mu_\mathrm{B}^2 B^2}{20\hbar^2}\sum_{j} \frac{\gamma_{\sing\trip_\pm}+1/\tau_{\mathrm{R},j}}{(\gamma_{\sing\trip_\pm}+1/\tau_{\mathrm{R},j})^2+(\epsilon_\sing-\epsilon_{\trip_\pm})^2} \sum_{b,b'}h_{j,b}h_{j,b'} {\Delta g_{b}^{(2)*}\Delta g_{b'}^{(2)} } \\
k_{\trip_0\trip_\pm}^{(\mathrm{g-aniso})} &= k_{\trip_\pm\trip_0}^{(\mathrm{g-aniso})} = \frac{\mu_\mathrm{B}^2 B^2}{5\hbar^2}\sum_{j} \frac{\gamma_{\trip_0\trip_\pm}+1/\tau_{\mathrm{R},j}}{(\gamma_{\trip_0\trip_\pm}+1/\tau_{\mathrm{R},j})^2+(\epsilon_{\trip_0}-\epsilon_{\trip_\pm})^2} \sum_{b,b'}h_{j,b}h_{j,b'} { \bar{g}_{b}^{(2)*}}\bar{g}_{b'}^{(2)} \\
k_{\trip_\mp\trip_\pm}^{(\mathrm{g-aniso})} &=k_{\trip_\pm\trip_\mp}^{(\mathrm{g-aniso})} = 0,
\end{align}
in which $\Delta g_{b}^{(2)} = g_{2,b}^{(2)} - g_{1,b}^{(2)}$ and $2\bar{g}_{b}^{(2)} = g_{1,b}^{(2)} + g_{2,b}^{(2)}$.
\section{$\text{PTZ}^{\bullet+}\text{--Ph}_\text{n}\text{--PDI}^{\bullet-}$ simulation parameters}
The following hyperfine constants were used in the simulation of the $\text{PTZ}^{\bullet+}\text{--Ph}_\text{n}\text{--PDI}^{\bullet-}$ magnetic field effects.\cite{Fay2017}
\begin{table}[h]
	$\text{PTZ}^{\bullet+}$
	\begin{tabular}{l|ccccccccc}
	$k$ & 1 & 2 & 3 & 4 & 5 & 6 & 7 & 8 & 9 \\
	\hline 
	Nucleus & H & H & H & H & H & H & H & H & N \\
	\hline
	$a_{k}/\text{mT}$ & $-$0.113 & $-$0.113 & $-$0.050 & $-$0.050 & $-$0.249 & $-$0.249 & 0.050 & 0.050 & 0.634	
	\end{tabular}\\
$\text{PDI}^{\bullet-}$
\begin{tabular}{l|cccccccc}
	$k$ & 1 & 2 & 3 & 4 & 5 & 6 & 7 & 8  \\
	\hline 
	Nucleus & H & H & H & H & H & H & N & N \\
	\hline
	$a_k$ & 0.0785 & 0.0785 & $-$0.1720 & $-$0.1720 & 0.0575 & 0.0575 & $-$0.0621 & $-$0.0621
\end{tabular}
\caption{Isotropic hyperfine coupling constants used in the $\text{PTZ}^{\bullet+}\text{--Ph}_\text{n}\text{--PDI}^{\bullet-}$ simulations.}
\end{table}

And the following electron spin system and recombination parameters were used.
\begin{table}[h]
	\begin{tabular}{l|cccc}
		Case & $k_\sing^\mathrm{r}/\text{ns}^{-1}$ & $k_\trip^\mathrm{r}/\text{ns}^{-1}$ & $2J/\text{mT}$ & $x$ \\
		\hline
		n=2 (original) & 0.04736 & 27.5 & $-$170 & 0.0416 \\
		n=3 (original) & 0.002160 & 3.8 & $-$31 & 0.525 \\
		n=3 (KME2 fit) & 0.002013 & 4.517 & $-$31 & 0.628\\
		n=4 (original) & 0.002449 & 0.35 & $-$6.4 & --\\
		n=4 (KME2 fit) & 0.001584 & 0.5322& $-$6.4 & --\\
		n=5 (original) & 0.002887 & 0.060 & $-$1.75 & --\\
		n=5 (KME2 fit) & 0.0009665 & 0.07533 & $-$1.47 & --
	\end{tabular}
	\caption{Rate constants, scalar coupling parameters and background corrections used in the simulations of the $\text{PTZ}^{\bullet+}\text{--Ph}_\text{n}\text{--PDI}^{\bullet-}$ radical pairs. Original parameters are taken from Ref. \onlinecite{Fay2017}.}
\end{table}

The background correction is a correction to the relative triplet yields of the form
\begin{align}
\text{RTY} = \frac{\Phi_\trip(B) + x}{\Phi_\trip(B=0) + x},
\end{align}
where the triplet yields are given by
\begin{align}
\Phi_\trip = k_\trip^\mathrm{r}\sum_{m=-1}^1\int_0^\infty \dd{t}  p_{\trip_{m}}(t).
\end{align}
The new fits were obtained by minimising the squared deviation from the experimental data in Ref. \onlinecite{Weiss2004} with the constraint that the radical pair lifetime, 
\begin{align}
\tau_{\mathrm{RP}} = \int_0^\infty \dd{t} \sum_{n} p_n(t),
\end{align}
is within 1 ns of the experimental lifetime.
\bibliography{kinetic-equations-v2}